\documentclass[smallextended]{svjour3}     
\smartqed  
\usepackage{graphicx}
\newcommand{\be}{\begin{equation}}
\newcommand{\ee}{\end{equation}}

\newcommand{\bea}{\begin{eqnarray}}
\newcommand{\eea}{\end{eqnarray}}

\newcommand{\dl}{\delta}

\newcommand{\ph}{\phi}

\newcommand{\rarrow}{\rightarrow}

 \journalname{General Relativity and Gravitation}
\begin{document}

\title{Charged cosmic strings interacting with gravitational and electromagnetic waves
} 

\titlerunning{Charged cosmic string interacting with...} 

\author{K Kleidis    \and
        A Kuiroukidis \and
        P Nerantzi     \and
        D B Papadopoulos}

\authorrunning{Kleidis, Kuiroukidis, Nerantzi, Papadopoulos} 
\institute{K. Kleidis \at
              Department of Physics,\\
              Aristotle University of Thessaloniki,\\
              GR-54124 Thessaloniki, Greece.\\
              Tel.: +302310998153\\
              Fax:  +302310995384\\
              \\
              \email{kleidis@astro.auth.gr}\\
              \\
              \emph{Present address:\\
              Department of Mechanical Engineering,\\
              Technological Education Institute of Serres,\\
              GR-62124 Serres, Greece.}                    
}
\date{Received: date / Accepted: date}

\maketitle

\begin{abstract}

Under a particular choice of the Ernst potential, we solve analytically the Einstein-Maxwell equations to derive {\em a new exact solution} depending on {\em five parameters}: the mass, the angular-momentum (per unit mass), $\alpha$, the electromagnetic-field strength, $k$, the parameter-$p$ and the Kerr-NUT parameter, $l$. This (Petrov Type D) solution is cylindrically-symmetric and represents the curved background around a charged, rotating cosmic string, surrounded by gravitational and electromagnetic waves, under the influence of the Kerr-NUT parameter. A C-energy study in the radiation zone suggests that both the incoming and the outgoing radiation is gravitational, strongly focused around the null direction and preserving its profile. In this case, the absence of the $k$-parameter from the C-energy implies that, away from the linear defect the electromagnetic field is too weak to contribute to the energy-content of the cylindrically-symmetric space-time under consideration. In order to explain this result, we have evaluated the Weyl and the Maxwell scalars near the axis of the linear defect and at the spatial infinity. Accordingly, we have found that the electromagnetic field is concentrated (mainly) in the vicinity of the axis, while falling-off prominently at large radial distances. However, as long as $k \neq 1$, the non-zero Kerr-NUT parameter enhances those scalars, both near the axis and at the spatial infinity, introducing some sort of {\em gravitomagnetic} contribution.



\end{abstract}

\section{Introduction}

\label{intro} Cosmic strings are one-dimensional objects that can be formed as linear defects at a symmetry-breaking phase-transition (for a detailed analysis see Hindmarsh and Kibble~\cite{r1} and/or Vilenkin and Shellard~\cite{r2}). If they exist, they may help us to explain some of the large-scale structures seen in the Universe today, such as gravitational lenses~\cite{r3},~\cite{r4}. They may also serve as {\em seeds} for density perturbations~\cite{r5}, as well as potential sources of relic gravitational radiation~\cite{r6}.

The curved space-time around a straight, isolated cosmic string is constructed by a flat background from which a wedge has been (locally) cut off. The resulting metric tensor acquires a {\em conical singularity} located on the axis of symmetry and the corresponding {\em angle-deficit} is given by $\dl \ph = 8 \pi \mu$, where $\mu$ is the {\em mass-density per unit length}~\cite{r1}. The cosmic-string radius is extremely small, of the order $10^{-27} \; m$~\cite{r7},~\cite{r8}. Hence, from the macroscopic point of view, a cosmic string represents an infinitely-long {\em line-source} and the gravitational field produced by it has {\em cylindrical symmetry}~\cite{r3}.

Cylindrically-symmetric solutions to the Einstein-Maxwell equations, pertinent to the linear defects produced by phase-transitions in the early Universe~\cite{r9},~\cite{r10}, have been modelled by several authors~\cite{r11} -~\cite{r23}. A particular method, developed by Xanthopoulos~\cite{r21} -~\cite{r23}, is based on the concept of the {\em Ernst potential} and the existing analogy between plane-waves and cylindrically-symmetric solutions of the Einstein
equations~\cite{r24}. Xanthopoulos~\cite{r22} derived a solution representing a rotating cosmic string surrounded by cylindrical gravitational waves, which (later) he extended, to include also electromagnetic waves~\cite{r23}. Garriga and Verdaguer~\cite{r25}, using the so called Belinski-Zakharov inverse scattering technique~\cite{r26}, \cite{r27}, have obtained several solutions describing straight cosmic strings interacting with solitonic-like gravitational waves. Based on the same method, Economou and Tsoubelis~\cite{r28} -~\cite{r30} derived a class of four-parameter, cylindrically-symmetric solutions to the Einstein equations in vacuum. On the other hand, Yazadjiev~\cite{r31}, \cite{r32} used a technique of generating solutions, which creates exact cosmic-string backgrounds from known solutions to the Einstein equations coupled to a massless scalar field. In this way, he managed to describe the curved space-time around a non-rotating cosmic string interacting with gravitational waves in the Einstein-Maxwell-dilaton gravity (EMDg). 

Recently, Bi\v{c}\'{a}k et al.~\cite{r33} (see also Lynden-Bell et al.~\cite{r34}) studied the linear and (mainly) the rotational {\em dragging effect}, that can be caused by a cylindrical gravitational wave on a local inertial frame. Adopting the Komar method~\cite{r35}, they have (thoroughly) analysed the concept of the angular momentum in cylindrically-symmetric space-times, demonstrating that it is non-vanishing only in the case where cylindrical gravitational waves {\em interact} with a rotating cosmic string~\cite{r33}. In this context, the metrics considered in~\cite{r22},~\cite{r23},~\cite{r28} and~\cite{r36}, do not possess angular momentum.

We see that there are many interesting solutions of the Einstein equations which can be generalized to include infinitely-long cosmic strings. In the present article, a linear topological defect is embedded in the so-called {\em Kerr-NUT space-time}~\cite{r24}. This is a class of axially symmetric space-times with an additional (the Newman-Unti-Tamburino) parameter $(l)$. In this context, the space-time surrounding the cosmic string is formed by removing a wedge of given deficit from the Kerr-NUT metric and then gluing together the resulting edges. Tomimatsu and Kihara~\cite{r37} and, more recently, Abdujabbarov et al.~\cite{r38} have raised several issues of astrophysical significance related to the NUT-parameter.

The Paper is organized as follows: In Section 2, we present some known aspects of the Einstein-Maxwell equations, based mainly on the work of Chandrasekhar and Xanthopoulos~\cite{r13} -~\cite{r15}. In Section 3, we use the method of Economou and Tsoubelis~\cite{r28},~\cite{r29}, to derive a new exact solution of the Einstein-Maxwell equations with non-zero Kerr-NUT parameter and in Section 4, we demonstrate that this solution describes a charged, rotating cosmic string surrounded by gravitational and electromagnetic waves.

\section{The Einstein-Maxwell equations}

\label{sec:1} Adopting the notation of Xanthopoulos~\cite{r13},~\cite{r22},~\cite{r23}, we consider a curved space-time with line-element that can be put in the form 
\begin{equation}
ds^2=e^{\nu+\mu_{3}}\sqrt{\Delta} \left [ \frac{d\eta^2}{\Delta}-\frac{d\mu^2}{\delta} \right ] -\frac{\Delta
\delta}{\Psi}d\phi^2-\Psi \left ( dz-q_{2}d\phi \right )^2 \; ,
\end{equation}
where we have set
\begin{equation}
\Delta=\eta^2+1 \; , \; \; \delta=\mu^2-1 
\end{equation}
(cf. Appendix A). The metric (1) admits two space-like, commuting Killing fields and the Einstein-Maxwell equations reduce to the Ernst equations~\cite{r24}
\begin{eqnarray}
&&\left ( ReZ-|H|^2 \right ) \left [ (\Delta Z_{,\eta})_{,\eta}-(\delta Z_{,\mu})_{,\mu} \right ]\nonumber\\
&=&\Delta(Z_{,\eta})^2-\delta(Z_{,\mu})^2-2H^{*} \left ( \Delta Z_{,\eta} H_{,\eta}-\delta Z_{,\mu} H_{,\mu} \right )
\end{eqnarray}
and
\begin{eqnarray}
&&\left ( ReZ-|H|^2 \right ) \left [ (\Delta H_{,\eta})_{,\eta}-(\delta H_{,\mu})_{,\mu} \right ] \nonumber\\
&=&\Delta H_{,\eta} Z_{,\eta}-\delta H_{,\mu}Z_{,\mu}-2H^{*} \left [ \Delta(H_{,\eta})^2-\delta(H_{,\mu})^2 \right ] \; ,
\end{eqnarray}
where
\begin{equation}
Z-HH^{*}=\Psi+i \Phi \; .
\end{equation}
In Eqs. (3) - (5), $\Psi$ and $\Phi$ are the Ernst potentials (e.g., see~\cite{r24}), $H$ measures the strength of the electromagnetic field and $H^{*}$ is the corresponding complex-conjugate quantity. In this gauge, for every solution to Eqs. (3) and (4), the metric coefficients $q_{2}$ and $e^{\nu+\mu_{3}}$ are obtained by the equations
\begin{equation}
q_{2,\eta}=\frac{\delta}{\Psi^2} \left [ \Phi_{,\mu}+2 Im(HH^{*}_{,\mu}) \right ] \; ,
\end{equation}
\begin{equation}
q_{2,\mu}=\frac{\Delta}{\Psi^2}\left [ \Phi_{,\eta}+2 Im(HH^{*}_{,\eta}) \right ]
\end{equation}
and
\begin{eqnarray}
\frac{\mu}{\delta}M_{,\eta}+\frac{\eta}{\Delta}M_{,\mu} &=& \frac{1}{\Psi^2} \left [ \Psi_{,\eta}\Psi_{,\mu} + (\Phi_{,\eta}+I_{(\eta)})(\Phi_{,\mu}+I_{(\mu)}) \right ] \nonumber \\ 
&+& \frac{2}{\Psi} \left ( H_{,\eta}H^{*}_{,\mu} + H^{*}_{,\eta}H_{,\mu} \right ) \; ,
\end{eqnarray}
\begin{eqnarray}
2\eta M_{,\eta}+2\mu M_{,\mu} &=& \left ( 4-\frac{3\eta^2}{\Delta}-\frac{\mu^2}{\delta} \right ) + \frac{4}{\Psi} \left ( \Delta H_{,\eta}H^{*}_{,n}+\delta H_{,\mu}H^{*}_{,\mu} \right ) \nonumber\\
&+&\frac{1}{\Psi^2}\left \lbrace \Delta \left [ \Psi^2_{,\eta}+(\Phi_{,\eta}+I_{(\eta)})^{2} \right ] + \delta \left [ \Psi^2_{,\mu} +(\Phi_{,\mu}+I_{(\mu)})^{2} \right ] \right \rbrace ,
\end{eqnarray}
where we have set
\begin{equation}
M=\nu+\mu_{3}+\ln\frac{\Psi}{\sqrt[4]{\Delta \delta}} \; , \; \;
I_{(a)}=2 Im(HH^{*}_{,a}) \; , \; \; a = \eta , \: \mu \; .
\end{equation}

\section{A new exact solution}

\label{sec:2}A first family of solutions to the Einstein-Maxwell equations describing an electrified cosmic string, can be derived by imposing the ansatz $Z = 1$. In this case, the corresponding metric is identical to that of Economou and Tsoubelis~\cite{r28} (see also~\cite{r29}) and it will not be considered any further. On the other hand, imposing the ansatz
\begin{equation}
H=Q(1+Z),~~k=(1-4QQ^{*})^{1/2} \; ,
\end{equation}
where $Q$ is a complex constant, a second family of electromagnetic string solutions can be obtained, as
\begin{equation}
ds^2=e^{\nu+\mu_{3}}_{(e)}\sqrt{\Delta} \left [ \frac{d\eta^2}{\Delta}-\frac{d\mu^2}{\delta} \right ] -\frac{\Delta
\delta}{\Psi_{(e)}}d\phi^2-\Psi_{(e)} \left ( dz-q_{2(e)}d\phi \right )^2
\end{equation}
We denote by $E$ the Ernst potential corresponding to the vacuum solution
\begin{equation}
\Psi+i\Phi=\frac{1+E}{1-E}
\end{equation}
of the Ernst equation
\begin{equation}
(1-EE^{*}) \left [ (\Delta E_{,\eta})_{,\eta}- (\delta E_{,\mu})_{,\mu} \right ] = -2E^{*} \left [ \Delta (E_{,\eta})^{2} - \delta (E_{,\mu})^{2} \right ] \; .
\end{equation}
Accordingly, we find
\begin{equation}
\Psi_{(e)}=\frac{k^2(1-EE^{*})}{|1-kE|^2} \: , ~~~
\Phi_{(e)}=\frac{ik(E^{*}-E)}{|1-kE|^2}
\end{equation}
and
\begin{equation}
Z_{(e)}=\frac{(1+kE)}{(1-kE)},~~k=(1-4QQ^{*})^{1/2},~~
H=\frac{2Q}{1-kE} \; ,
\end{equation}
so that
\begin{equation}
e_{(e)}^{\nu+\mu_{3}}=\frac{1}{4k^2} \left [ (1-k)^2 \left ( \Psi_{(\upsilon)}^2 + \Phi_{(\upsilon)}^2 \right ) + 2 (1-k^2)\Psi_{(\upsilon)}+(1+k^2) \right ] e_{(\upsilon)}^{\nu+\mu_{3}} \; ,
\end{equation}
where the subscripts $(e)$ and $(\upsilon)$ stand for the electric and the vacuum solution, respectively. As regards $q_{2(e)}$, it is obtained from the equation
\begin{equation}
q_{2(e)}=\frac{(1+k)^2}{4k^2}q_{2(\upsilon)}+\frac{(1-k)^2}{4k^2}q_2^{(e)} \; ,
\end{equation}
where, $q_{2(\upsilon)}$ is given by Eq. (A23) of the Appendix A and $q_2^{(e)}$ is a solution to the system of differential equations
\begin{equation}
q^{(e)}_{2,\eta}=\frac{\delta}{\Psi^2_{(\upsilon)}} \left [ \left ( \Phi^2_{(\upsilon)} - \Psi^2_{(\upsilon)} \right ) \Phi_{(\upsilon),\mu} + 2\Phi_{(\upsilon)}\Psi_{(\upsilon)}\Psi_{(\upsilon),\mu} \right ]
\end{equation}
and
\begin{equation}
q^{(e)}_{2,\mu}=\frac{\Delta}{\Psi^2_{(\upsilon)}} \left [ \left ( \Phi^2_{(\upsilon)} - \Psi^2_{(\upsilon)} \right ) \Phi_{(\upsilon),\eta} + 2\Phi_{(\upsilon)}\Psi_{(\upsilon)}\Psi_{(\upsilon),\eta} \right ] \; ,
\end{equation}
where $\Psi^2_{(\upsilon)}$ and $\Phi^2_{(\upsilon)}$ are given by Eqs. (A24) and (A26) of the Appendix A, respectively. Demanding that Eqs. (19) and (20) satisfy the integrability condition $q_{2, \eta \mu}^{(e)} = q_{2, \mu \eta}^{(e)}$, we find 
\begin{equation}
q_2^{(e)}=\frac{2}{pY} \left [ q\delta(1+p\eta)+lq\delta(l+q)-p^2\Delta(\mu-1) \right ] \; ,
\end{equation}
where the resulting integration constant is suitably chosen so that $q_2^{(e)}$ vanishes for $\mu = 1$ (on the azimuthal axis). Since the metric coefficient $q_2^{(e)}$ is determined up to an additive constant, we use this freedom to simplify the expression of $q_2^{(e)}$. Eventually, Eqs. (15) - (21), with the aid of Eq. (A23), give us all the   coefficients of the metric (12), in the form
\begin{eqnarray}
q_{2(e)}&=&\frac{(1+k)^2}{2k^2 p Y} \left [ q\delta(1-\eta p)+q\delta l(l-q)+lp^2(\mu-1)\Delta \right ] \nonumber\\
&+&\frac{(1-k)^2}{2k^2 p Y} \left [ q\delta(1+p\eta)+q\delta l(l+q)-lp^2(\mu-1)\Delta \right ] \; , 
\end{eqnarray}
\begin{equation}
\Psi_{(e)}=k^2\frac{Y}{\Pi} \: ,~~~\Phi_{(e)}=2k\frac{(q\mu-lp\eta)}{\Pi} \; , 
\end{equation}
and
\begin{equation}
e_{(e)}^{\nu+\mu_{3}}\sqrt{\Delta}=\frac{\alpha^2\Pi}{k^2}
\end{equation}
with $q=\sqrt{(1+p^2+l^2)}$, $\Pi=(k-p\eta)^2+(q\mu-k l)^2$ and $Y=p^2\Delta+q^2\delta$. In accordance, the line element (12) reads 
\begin{equation}\label{m1}
ds^2=\frac{\alpha^2 \Pi}{k^2} \left [ \frac{d \eta^2}{\Delta}-\frac{d\mu^2}{\delta} \right ]-\frac{\Delta\delta \Pi}
{k^2Y} d\phi^2-\frac{k^2Y}{\Pi} \left [ dz-q_{2(e)}d\phi \right ]^2
\end{equation}
The metric (\ref{m1}) satisfies the Rainich-Wheeler-Misner conditions (e.g., see p. 529 of \cite{r39}) and, therefore, is {\em a new exact solution} to the Einstein-Maxwell equations, depending on {\em five parameters} (although the form of Eq. (25) does not render their physical interpretation quite clear): the mass and the angular-momentum per unit mass, $\alpha$ (in fact, both parameters are used in a way that reflects their physical meaning in the context of the original stationary, axially-symmetric space-time), the strength of the electromagnetic-field, $k$, the parameter-$p$ and the Kerr-NUT parameter, $l$. Based on the results of Bi\v{c}\'{a}k et al.~\cite{r33}, the fact that, in our case, $\alpha \neq 0$ (in contrast to~\cite{r22},~\cite{r23} and~\cite{r28}), indicates that the Kerr-NUT parameter $(l)$ introduces some sort of {\em coupling} between the cylindrical gravitational waves and the rotating cosmic string. A potential source of this coupling could be the magnetic properties that (should be) attributed to the charged linear defect due to its rotation. Indeed, it has been recently recognized that $l$ corresponds to the {\em gravitomagnetic monopole moment} of a charged Kerr-NUT compact object~\cite{r38}. Whether this aspect applies also in our case is an open question, the answer of which will be the scope of a future work.

At present, we can conclude that, the solution (25) describes the space-time around a charged $(k \neq 1)$, rotating $(\alpha \neq 0)$ linear defect, surrounded by gravitational and electromagnetic waves. In fact, the ansatz (11) - formely imposed to construct the Kerr-Newman solution from the corresponding Kerr one (e.g., see~\cite{r24}) - has been used to generate solutions of the Einstein-Maxwell equations, describing the curved space-time which results from the collision of gravitational and electromagnetic waves (e.g., see~\cite{r14},~\cite{r15}).

Upon consideration of the metric (\ref{m1}), a convenient Newman-Penrose null basis may be found as in Xanthopoulos~\cite{r22},~\cite{r23}. The contravariant components of this basis are written in the form
\begin{equation}
l^{\mu} = \left ( \frac{1}{\sqrt{2}}\frac{k\sqrt{\eta^{2}+\mu^{2}}}{\alpha \sqrt{\Pi}} , \frac{1}{\sqrt{2}} \frac{k\sqrt{\eta^{2}+\mu^{2}}}{\alpha \sqrt{\Pi}}, \: 0, \: 0 \right ) \; ,
\end{equation}
\begin{equation}
n^{\mu} = \left ( \frac{1}{\sqrt{2}} \frac{k\sqrt{\eta^{2}+\mu^{2}}}{\alpha \sqrt{\Pi}} , - \frac{1}{\sqrt{2}} \frac{k\sqrt{\eta^{2}+\mu^{2}}}{\alpha \sqrt{\Pi}}, 0, 0 \right ) \; ,
\end{equation}
\begin{equation}
m^{\mu} = \left ( 0, \: 0, \: \frac{1}{\sqrt{2}} i \frac{k\sqrt{Y}}{\omega\sqrt{\Pi}}, \frac{1}{\sqrt{2}} \frac{ik^{2}q_{2(e)}Y + \omega \Pi}{\omega k \sqrt{\Pi Y}} \right ) \; \; \;
\end{equation}
and
\begin{equation}
\bar{m}^{\mu} = \left ( 0, 0, - \frac{1}{\sqrt{2}} i \frac{k\sqrt{Y}}{\omega\sqrt{\Pi}}, \frac{1}{\sqrt{2}} \frac{-ik^{2}q_{2(e)}Y + \omega \Pi}{\omega k \sqrt{\Pi Y}} \right ) \; .
\end{equation}
In fact, one may use the above vectors to reconstruct the metric (\ref{m1}) by making use of the relation $g_{\mu\nu} = l_{\mu}n_{\nu}+l_{\nu}n_{\mu}-m_{\mu}\bar{m}_{\nu}-m_{\nu}\bar{m}_{\mu}$. Furthermore, the vectors (26) - (29) satisfy the orthogonality conditions $l\cdot m=l\cdot\bar{m}=n\cdot m=n\cdot\bar{m}=0$, as well as the requirements $l\cdot l = n \cdot n=m\cdot m=\bar{m}\cdot \bar{m}=0$ (in other words, they are {\em null}) and the normalization conditions $l\cdot n = 1$ and $m\cdot\bar{m}=-1$~\cite{r40},~\cite{r41}.

Using the above null tetrad, the only non-vanishing Weyl and Maxwell scalars for the metric (\ref{m1}) are given by the equations (3.43a) - (3.43f) of Xanthopoulos~\cite{r23} and they are expressed in terms of the Ernst potential $E$ (arising from the potentials $\Psi$ and $\Phi$)
\begin{equation}
E = \frac{1-il}{p\eta-iq\mu} \; .
\end{equation}
Accordingly, the Weyl scalars read
\begin{equation}
\Psi_{2}=-\frac{k^{2}}{2\alpha^{2}}\frac{(1-il)^{2}}{1+l^{2}}\frac{(E-k)(E^{*})^{3}}{(1-kE^{*})^{3}(1-kE)} \; ,
\end{equation}
\begin{equation}
\frac{1-\cal{E}}{1-\cal{E^*}}\Psi_{0}=\frac{3k^{2}}{2\alpha^{2}}\frac{E(E-k)(\sqrt{\Delta}E^{*}_{,\eta}+\sqrt{\delta}E^{*}_{,\mu})^{2}}{(1-EE^{*})(1-kE)(1-kE^{*})^{3}} \; ,
\end{equation}
\begin{equation}
\frac{1-\cal{E^*}}{1-\cal{E}}\Psi_{4}=\frac{3k^{2}}{2\alpha^{2}}\frac{E(E-k)(\sqrt{\Delta}E^{*}_{,\eta}- \sqrt{\delta}E^{*}_{,\mu})^{2}}{(1-EE^{*})(1-kE)(1-kE^{*})^{3}} \; ,
\end{equation}
while, the corresponding Maxwell ones are written in the form
\begin{equation}
\Phi_{00} = \frac{k^{2}(1-k^{2})}{2\alpha^{2}}\frac{(EE^{*})^{2}}{(1-kE)^{2}(1-kE^{*})^{2}} = \Phi_{22} 
\end{equation}
and
\begin{equation}
\frac{1-\cal{E^*}}{1-\cal{E}}\Phi_{20}=-\frac{k^{2}(1-k^{2})}{2\alpha^{2}}\frac{EE^{*}(\sqrt{\Delta}E_{,\eta}+\sqrt{\delta}E_{,\mu})(\sqrt{\Delta}E^{*}_{,\eta}-\sqrt{\delta}E^{*}_{,\mu})}{(1-EE^{*})(1-kE)^{2}(1-kE^{*})^{2}} \; .
\end{equation}
For $l=0$, all the above equations reduce to the corresponding formulas predicted by Xanthopoulos~\cite{r23}. For $l \neq 0$, we verify that the consistency condition for the Maxwell scalars
\begin{equation}
\Phi_{20}\cdot\Phi_{20}^{*}=\Phi_{00}\cdot\Phi_{22}
\end{equation}
is satisfied, while, their combination to the Weyl scalars satisfies the identities
\begin{equation}
\Psi_{0}\cdot\Psi_{4}=9\Psi_{2}^{2}
\end{equation}
and
\begin{equation}
3\Phi_{00}\cdot\Psi_{2}=\Phi_{20}\cdot\Psi_{0} \; ,
\end{equation}
which are the necessary and sufficient conditions for a metric to be of Petrov Type D and the twice-repeated principal null directions of the Weyl and Maxwell tensors to coincide~\cite{r40},~\cite{r41}.

\section{The electrified cosmic string at $\omega\rightarrow 0^{+}$, $\omega\rightarrow\infty$}

\label{sec:4}In what follows, we discuss both the mathematical and the physical properties of the space-time (\ref{m1}), together with a measure of its energy-content (Thorne's C-energy~\cite{r42}) and the Weyl scalars in three characteristic regions of the problem: (i) Near the axis of the linear defect, (ii) at the spatial infinity and (iii) at the vicinity of the null cone.

\subsection{The metric at $\omega\rightarrow 0^{+}$, $\omega\rightarrow\infty$}

\label{sec:5} To investigate the behavior of the metric coefficients $q_{2(e)}$, $e_{(e)}^{\nu+\mu_{3}}$ and $\Psi_{(e)}$ near the axis and at spatial infinity, we express the metric (\ref{m1}) in cylindrical coordinates using Eqs. (A15) and (A16) (e.g., see~\cite{r18},~\cite{r19} and~\cite{r22},~\cite{r23}). The expressions of the metric coefficients in terms of $t \in \Re$ and $\omega \in [0,\infty)$ are quite difficult to handle. We can bypass this problem, by considering the Taylor expansion of $(\eta, \: \mu)$ in terms of $(t, \: \omega)$. The corresponding results also consist of lengthy expressions, which (however) can be simplified (considerably) in the limits $\omega \rightarrow 0^{+}$ and $\omega\rightarrow\infty$. Accordingly, {\em near the axis} $(\omega \rightarrow 0^{+})$ we have
\begin{equation} 
\eta=t-\frac{\omega^2 t}{2(1+t^2)}+O(\omega^4),~~\mu=1+\frac{\omega^2}{2(1+t^2)}+O(\omega^4)
\end{equation}
and, hence, for $\omega \ll t$, we obtain
\begin{equation}
\frac{\alpha^2 \Pi}{k^2(\eta^2+\mu^2)}\sim \frac{\alpha^2 N}{k^2},~~\frac{\omega^2\Pi}{k^2 Y}\sim \frac{\omega^2 N}{k^2p^2}
\end{equation}
and
\begin{equation}
\frac{k^2 Y}{\Pi}\sim \frac{k^2p^2}{N},~~q_{2(e)}\sim\frac{\omega^2 B}{k^2p^3(1+t^2)^2}
\end{equation}
where, we have set
$$B = q (1+k^2-2kpt)+q l^2 \left ( 1+k^2 \right )+ kl \left [ -2q^2+p^2 \left ( 1+t^2 \right ) \right ]$$
and
$$N = \frac{(k - tp)^2+(q-kl)^2}{1+t^2} \: .$$
Consequently, near the axis, the line-element reads
\begin{equation}\label{s0}
ds^2=\frac{\alpha^2N}{k^2} \left [ dt^2-d\omega^2-\frac{\omega^2}{\alpha^2 p^2} d\phi^2 \right ] - \frac{k^2p^2}{N} \left [ dz-\frac{\omega^2 B}{k^2 \left ( 1+t^2 \right )^2 p^3} d\phi \right ]^2 \; .
\end{equation}
It is evident that the corresponding curvature is {\em smooth} and the measures of the two Killing vectors are $\mid\frac{\partial }{\partial z}\mid^2=O(1)$ and $\mid\frac{\partial }{\partial \phi}\mid^2=O(\omega^2)$, which are first-order orthogonal, that is $\frac{\partial }{\partial z}\frac{\partial }{\partial \phi}=O(\omega^2)$. Eq. (\ref{s0}) implies that, given a small circle lying on the hypersurface $dt=0=dz$ and having its center at $\omega=0$, the ratio circumference/radius differs from $2\pi$, unless $\mid \alpha p \mid=1$. When $\mid \alpha p \mid\neq 1$ the region near the symmetry axis is characterized by an {\em angle-deficit} and the metric (\ref{s0}) exhibits a {\em conical singularity}. In particular, the angle-deficit around this linear defect is given by~\cite{r31},~\cite{r32}
\begin{equation}
\delta\phi_{axis}=2\pi-\lim_{\rho\rightarrow0}\frac{\int_0^{2\pi}\sqrt{g_{\phi\phi}}d\phi}{\int_0^{\rho}\sqrt{g_{\rho\rho}}d\rho} = 2\pi \left [ 1 - \frac{1}{\mid \alpha p\mid} \right ]
\end{equation}
and the corresponding mass-density is $\mu_0=\frac{1}{4} \left [ 1-\frac{1}{\mid \alpha p\mid} \right ]$. In this case, the Kerr-NUT parameter does not make any contribution to the problem and the result is the same as in the $l=0$ case.

On the other hand, {\em away from the axis} $(\omega \gg t)$, we have
\begin{equation}
\eta=\frac{t}{\omega}+\frac{t(t^2-1)}{2\omega^3}+O(\omega^{-4}),~~\mu=\omega+\frac{(1-t^2)}{2\omega}+O(\omega^{-2})\end{equation}
\begin{eqnarray}
\frac{\alpha^2 \Pi}{k^2(\eta^2+\mu^2)} & = & \frac{\alpha^2 q^2}{k^2}-\frac{2\alpha^2 lq}{k\omega} + O(\omega^{-2}) \;, \nonumber\\
\frac{\omega^2\Pi}{k^2 Y} & = & \frac{\omega^2}{k^2}-\frac{2 l \omega}{k q} + \frac{(1+l^2)(1+k^2)}{k^2q^2} + O(\omega^{-3}) \; , \\
\frac{k^2 Y}{\Pi} & = & k^2+\frac{2lk^3}{q\omega}+O(\omega^{-2}) \nonumber
\end{eqnarray}
and
\begin{equation}
q_{2(e)}=\frac{1}{k^2 pq}\Lambda+O(\omega^{-1}) \; ,
\end{equation}
where we have set $\Lambda= \left [ (1+k^2)(1+\frac{l^2}{2})+kl(l-2q) \right ]$. In accordance, away from the axis $(\omega \rightarrow \infty)$, the line-element (\ref{m1}) reads
\begin{eqnarray}\label{s1}
ds^2 & = & \frac{\alpha^2q^2}{k^2} \left [ 1-\frac{2kl}{q\omega}+O(\omega^{-2}) \right ] \left [ dt^2 - d \omega^2 - \frac{\omega^2}{\alpha^2 q^2} d\phi^2 \right ]\nonumber\\
&-&k^2 \left [ 1+\frac{2kl}{q\omega}+O(\omega^{-2}) \right ] \left [ dz-\frac{1}{qpk^2}\Lambda d\phi \right ]^2 \; .
\end{eqnarray}
For the metric (\ref{s1}), the angle-deficit is given by the formula,
\begin{equation}\label{s1a}
\delta\phi_{\infty} = 2\pi-\lim_{\rho\rightarrow \infty} \frac{\int_0^{2\pi}\sqrt{g_{\phi\phi}}d\phi}{\int_0^{\rho}\sqrt{g_{\rho\rho}}d\rho}
= 2\pi \left [ 1 - \frac{1}{\mid\alpha q\mid} \right ] \; .
\end{equation}
Eq. (\ref{s1a}) implies that the electromagnetic field does not make any contribution to the angle-deficit. In this case, the mass-density of the linear defect is given by the equation
\begin{equation}
\mu_0=\frac{1}{4} \left [ 1 - \frac{1}{\mid\alpha q\mid} \right ] \; .
\end{equation}
As regards the metric (\ref{s1}), the measures of the two Killing vectors are $\mid\frac{\partial }{\partial z}\mid^2=O(1)$ and $\mid\frac{\partial }{\partial \phi}\mid^2=O(\omega^2)$. These are not hypersurface orthogonal [not even in the first order since $\frac{\partial }{\partial z}\frac{\partial }{\partial \phi}=O(\omega)$] unless $l=0$ or $k=1$. Combining  Eqs. (42) and (48) we find that
\begin{equation}\label{def1}
\delta \phi_{asym} >\delta \phi_{axis} \; ,
\end{equation}
since $q^2=1+p^2+l^2$. In other words, the angle-deficit as measured asymptotically is always greater than the corresponding deficit as measured near the axis. This excess in the deficit is (probably) attributed to the contribution of the energy of the intervening gravitational waves. Thus the choice $\mid \alpha \mid=q^{-1}$, which would erase the asymptotic deficit, requires a string with negative mass-density. Therefore, although near the axis the string can be erased for a suitable choice of the parameter $\alpha$ (e.g., $\mid \alpha\mid=p^{-1}$), the asymptotic angle-deficit can not be eliminated in a physically acceptable
situation.

\subsection{The C-energy}

\label{sec:6} As regards the cylindrically-symmetric solutions to the Einstein equations, the quantity
\begin{equation}\label{cen}
C=\nu+\frac{1}{2} \ln \left [ \Psi_{(e)} \right ]
\end{equation}
is often referred to as their {\em C-energy}~\cite{r42}, being proportional to the energy-density per unit length contained in a cylinder of radius $\omega$. However, recently, Ashtekar et al.~\cite{r43},~\cite{r44}, demonstrated that Thorne's C-energy gives the correct energy expression for cylindrically-symmetric space-times only in the {\em weak-field} limit. With this in mind, we calculate the C-energy for the metric (25). In this case, with the aid of Eqs. (23) and (24), Eq. (\ref{cen}) is written in the form
\begin{equation}
C=\frac{1}{2}\ln \left [ \frac{\alpha^2 Y}{(\eta^2+\mu^2)} \right ] \; .
\end{equation}
Notice that Eq. (52) does not depend on $k$. This feature of the C-energy implies that, away from the linear defect the electromagnetic field is too weak to contribute to the energy-content of the space-time (25) [as it happens also with the case of the angle-deficit at infinity (cf. Eq. (48)]. A physical explanation of this result can be obtained in terms of the Weyl and the Maxwell scalars and is presented in the next Section. Accordingly, we expect that, both the incoming and the outgoing radiation consists solely of gravitational waves. It is easy for someone to see that, near the axis (i.e., as $\omega \rightarrow 0$) $C_{axis}\sim \ln \mid \alpha p\mid$, while, asymptotically (as $\omega\rightarrow \infty$), $C_{asym}\sim \ln \mid \alpha q \mid$. Since $q^2 = 1+p^2+l^2$, we have $C_{asym} > C_{axis}$. This result could be a manifestation that the intervening gravitational waves contribute to the curved space-time a positive energy-amount. Furthermore, the behavior of the C-energy flux along the null-directions reveals that the presently considered solution, although it is quite tedious, exhibits the same radiative behavior and the same fall-off properties away from the null-direction of propagation as the solution of Economou and Tsoubelis~\cite{r28} -~\cite{r30}.

To examine its behavior, we introduce the so-called {\em retarded} and {\em advanced} null-coordinates, $u=t-\omega$ and $v=t+\omega$, for the metric (\ref{m1}) and we express the C-energy and its first derivatives $\partial C / \partial u = C_{,u}$ and $\partial C / \partial v = C_{,v}$ in terms of $u$ and $v$. Accordingly, we obtain:
\\

{\em Behavior at past null-infinity:}
\begin{eqnarray}
\lim_{u\rightarrow -\infty}C_{,u} & = & O(1/u^2) \; , \\
\lim_{u\rightarrow -\infty}C_{,v} & = & \frac{1+l^2}{[(p^2+q^2)\sqrt{(1+v^2)}+v(1+l^2)](1+v^2)}+O(1/u) \; . \nonumber
\end{eqnarray}

{\em Behavior at future null-infinity:}
\begin{eqnarray}
\lim_{v\rightarrow \infty}C_{,u} & = & -\frac{1+l^2}{[(p^2+q^2)\sqrt{(1+u^2)}-u(1+l^2)](1+u^2)}+O(1/v) \;, \nonumber \\ \lim_{v\rightarrow \infty}C_{,v} & = & + O(1/v^2) \; .
\end{eqnarray}
We observe that, at the past null-infinity there is a flux of incoming gravitational radiation toward the axis, with its profile given by Eq. (53). At large values of $v$, Eq. (53) behaves as $\lim_{u \rightarrow - \infty} C_{,v}\sim \frac{1+l^2}{(1+l^2+q^2) v^3}$. Thus, the original pulse of incoming radiation is concentrated around $v=0$ and for large values of $v$ it falls-off quite rapidly, i.e., as $v^{-3}$. Therefore, it may interpreted that, near the past null-infinity there is a beam of incoming radiation. Similarly, at large values of $u$, Eq. (54) behaves as $\lim_{u\rightarrow \infty} C_{,u}\sim - \frac{1+l^2}{p^2 u^3}$, suggesting that, near the future null-infinity, there is only outgoing null radiation, which is beamed around $u=0$.

These results indicate that the entire space-time (\ref{m1}) is filled with a mixture of incoming and outgoing gravitational radiation, originated at the past null-infinity, which is reflected by the cosmic string and propagate toward the future null-infinity. Both waves, incoming and outgoing, are beamed around the null-direction in the radiation zone.

\subsection{The Weyl and Maxwell scalars close to the axis and at infinity}

\label{sec:7} Accordingly, we evaluate the Weyl and the Maxwell scalars corresponding to Eq. (\ref{m1}). The expressions are quite long to be of any physical significance. Therefore, we are interested only in their asymptotic behavior near the axis, $\omega \rarrow 0$ and in the asymptotic region $\omega \rarrow \infty$.

{\em In the vicinity of the axis}, with the aid of Eqs. (31) - (35), we find that all the Weyl and Maxwell scalars approach finite, non-zero values, which do not depend on time, but only on the parameters $\alpha$, $k$ and $l$. In particular, the Weyl scalars read
\begin{equation}
\Psi_{2}\rightarrow-\frac{k^{2}}{2\alpha^{2}}(1+l^{2})(1+il)\frac{1+i(kq-l)}{[k^2+(q-kl)^{2}][-k+i(q-kl)]^{2}} \; ,
\end{equation}
\begin{equation}
\frac{1-\cal{E}}{1-\cal{E^*}} \Psi_{0} \rightarrow \frac{3k^{2}}{2\alpha^{2}} (1+l^{2}) (1+il) \frac{1+i(kq-l)}{[k^2+(q-kl)^{2}][-k+i(q-kl)]^{2}} 
\end{equation}
and
\begin{equation}
\frac{1-\cal{E^*}}{1-\cal{E}}\Psi_{4}\rightarrow\frac{3k^{2}}{2\alpha^{2}}(1+l^{2})(1+il)\frac{1+i(kq-l)}{[k^2+(q-kl)^{2}][-k+i(q-kl)]^{2}} \; ,
\end{equation}
while, the Maxwell ones are written in the form
\begin{equation}
\Phi_{00} \rightarrow \frac{k^{2}(1-k^{2})}{2\alpha^{2}}(1+l^{2})^{2}\frac{1}{[k^2+(q-kl)^{2}]^{2}} \leftarrow \Phi_{22}
\end{equation}
and
\begin{equation}
\frac{1-\cal{E^*}}{1-\cal{E}}\Phi_{20}\rightarrow-\frac{k^{2}(1-k^{2})}{2\alpha^{2}}(1+l^{2})^{2}\frac{1}{[k^2+(q-kl)^{2}]^{2}} \; .
\end{equation}
We observe that the presence of a non-zero Kerr-NUT parameter results in the enhancement of both the Weyl (as $l^2$) and the Maxwell (as $l^4$) scalars. 

{\em At the spatial infinity} (i.e., as $\omega \rarrow \infty$), the Weyl scalars behave as
\begin{equation}
\Psi_{2} = \frac{ik^{3}}{2\alpha^{2}q^{3}\omega^{3}}(1+l^{2})(1+il)+O(\omega^{-4}) \; ,
\end{equation}
\begin{equation}
\frac{1-\cal{E}}{1-\cal{E^*}} \Psi_{0} = \frac{3ik^{3}}{2\alpha^{2}q^{3}\omega^{3}}(1+l^{2})(1+il)+O(\omega^{-4})
\end{equation}
and
\begin{equation}
\frac{1-\cal{E^*}}{1-\cal{E}}\Psi_{4} = \frac{3ik^{3}}{2\alpha^{2}q^{3}\omega^{3}}(1+l^{2})(1+il)+O(\omega^{-4}) \; ,
\end{equation}
while, the Maxwell ones as
\begin{equation}
\Phi_{00} = \frac{k^{2}(1-k^{2})}{2\alpha^{2}q^{4}\omega^{4}}(1+l^{2})^{2}+O(\omega^{-5}) = \Phi_{22}
\end{equation}
and
\begin{equation}
\frac{1-\cal{E^*}}{1-\cal{E}}\Phi_{20} = \frac{k^{2}(1-k^{2})}{2\alpha^{2}q^{4}\omega^{4}}(1+l^{2})^{2} + O(\omega^{-5}) \; .
\end{equation}
In this case, we observe that the Weyl scalars fall-off as $\omega^{-3}$, while the Maxwell ones as $\omega^{-4}$. The latter result suggests that, as we depart from the linear defect, the electromagnetic field decreases quite prominently, thus becoming insignificant at large radial distances; probably too weak to contribute to the C-energy. On the other hand, both the Weyl and the Maxwell scalars are (once again) enhanced due to the non-zero Kerr-NUT parameter (also as $l^2$ and $l^4$, respectively), indicating that $l$ can {\em mediate} in an interaction between gravitational waves and a charged, rotating linear defect, probably, by the introduction of a {\em gravitomagnetic} contribution (in connection, see~\cite{r38}).

\section{Discussion}

\label{sec:8} In the present article, we have solved analytically the Einstein-Maxwell equations with the particular choice of the Ernst potential $E=\frac{1-il}{E_k^{*}}$, where $E_k = p\eta+iq\mu$ is the corresponding potential of the Kerr solution. Accordingly, we have derived {\em a new exact solution} to the Einstein-Maxwell equations depending on {\em five} parameters: the mass, the (specific) angular-momentum per unit mass $(\alpha)$, the electromagnetic-field strength $(k)$, the parameter-$p$ and the Kerr-NUT parameter $(l)$. The (Petrov Type D) solution (\ref{m1}) has cylindrical symmetry and represents the curved background around a charged $(k \neq 1)$, rotating $(\alpha \neq 0)$ cosmic string, surrounded by gravitational and electromagnetic waves under the influence of the Kerr-NUT parameter. In what follows, we summarize the most important mathematical and/or physical properties of this space-time.

In the absence of the electromagnetic field (i.e., for $k = 1$), the line-element (\ref{m1}) is reduced to the corresponding solution obtained by Economou and Tsoubelis~\cite{r28} -~\cite{r30}, while for $k = 1$ and $l = 0$ it results in the solution obtained by Xanthopoulos~\cite{r22}, describing a cosmic string of zero angular momentum in the presence of cylindrically-symmetric gravitational waves. On the other hand, for $l = 0$ and $k \neq 1$ one finds the solution obtained also by Xanthopoulos~\cite{r23}, regarding a linear defect in the space-time of colliding gravitational and electromagnetic waves, which (once again) do not interact with the cosmic string, since, in this case, $\alpha = 0$, as well (in connection, see~\cite{r33}).

In our case (i.e., for $l \neq 0$ and $k \neq 1$), the space-time is smooth everywhere and exhibits a conical singularity, both near the axis and at the infinity. The angle-deficit near the axis is always smaller than the corresponding deficit as determined asymptotically. The deficit near the axis signals the existence of a cosmic string with mass per unit length given by Eq. (43). On the other hand, the deficit at infinity is attributed to the combined effect of the string and of the energy carried by gravitational radiation. The particular choice $\mid \alpha \mid = p^{-1}$ erases the string at small distances. In a similar fashion, one could erase also the asymptotic deficit, but this would require a linear defect with negative mass per unit length. Therefore, the angle-deficit at infinity cannot be erased.

The study of the C-energy in the radiation zone suggests that both the incoming and the outgoing radiation is gravitational, very strongly focused around the null-direction and preserves its profile. The absence of the $k$-parameter from the C-energy and its derivatives suggests that, away from the linear defect the electromagnetic field is too weak to contribute to the energy-density of the cylindrically-symmetric space-time (25). For the same reason, there is no electromagnetic flux either near or far from the linear defect. In order to explain these results, we have evaluated the (Weyl and the) Maxwell scalars near the axis of the linear defect and at the spatial infinity. Accordingly, we have found that the electromagnetic field is concentrated (mainly) in the vicinity of the axis, while falling-off prominently (as $\omega^{-4}$) at large radial distances. However, as long as $k \neq 1$, the non-zero Kerr-NUT parameter enhances the Maxwell scalars both near the axis and at infinity (cf. Eqs. (58), (59) and Eqs. (63), (64), respectively), introducing some sort of a {\em gravitomagnetic} contribution~\cite{r38}. 

Other remarkable features of the space-time (\ref{m1}) in the vicinity of the axis, are:

1. For $\mid \alpha p \mid > 1$, i.e., at time-like infinity, the axis region of the metric (\ref{s0}) is flat with a conical singularity and the axis is occupied by a static string.

2. The original Killing vectors $\frac{\partial}{\partial z}$ and $\frac{\partial}{\partial \phi}$ of the metric (\ref{m1}) are not hypersurface orthogonal at large distances (although they are in the vicinity of the axis). This is (probably) due to the non-zero value of the Kerr-NUT parameter and the strength of the electromagnetic field. However, with a transformation of the form $z\rightarrow \tilde{z}=z-\Lambda \phi$, where $\Lambda = \frac{1}{qpk^2} \left [ (1+k^2) + \frac{l^2}{2}(1+k^2) + kl(l-2q) \right ]$, the last term in Eq. (\ref{s1}) can be gauged away (only locally) and the corresponding Killing vectors $\frac{\partial}{\partial \tilde{z}}, \frac{\partial}{\partial \phi}$ can become hypersurface orthogonal. In any other case, the last term of the metric (\ref{s1}) will give rise to global effects,
analogous to those arising in stationary space-times, when the time-like Killing vector is not hypersurface orthogonal~\cite{r28}.

We believe that the metric (\ref{m1}) might be useful to discuss some astrophysical implications, either close to the string or far away from the electrified linear defect, in the spirit of Abdujabbarov et al.~\cite{r38} and/or Hiscock~\cite{r45}. In particular, in a forthcoming paper we intend to examine MHD phenomena in the frame of the metric (\ref{m1}), which may lead to stability criteria of astrophysical significance.

\section {Appendix A}

\label{sec:9}For completeness, we review the techniques of generating solutions to the Einstein-Maxwell equations with
cylindrical symmetry, as formulated by Chandrasekhar and Xanthopoulos~\cite{r13} -~\cite{r15}. The corresponding line-element for a vacuum space-time reads
$$ds^2 = e^{2\nu} \left [ dt^2-d\omega^2 \right ]-\frac{\omega}{\chi} \left [ dz-q_{2} d\phi \right ]^2-\omega \chi d\phi^2 \; , \eqno{(A1)}$$
where $\partial/\partial z$ and $\partial/\partial \phi$ are the axial and the azimuthal Killing fields, while $\nu, \; \chi,$ and $q_2$ depend only on $t$ and $\omega$. The Weyl and Maxwell scalars of a solution to the Einstein-Maxwell equations may be derived both in terms of the Ernst potential denoted by ${\cal E}$ (arising from the metric coefficients $\chi$ and $q_2$) or by $E$ (arising from the metric coefficients $\Phi$ and $\Psi$). Setting
$$\chi+iq_2={\cal Z}=\frac{1+{\cal E}}{1-{\cal E}} \; , \eqno{(A2)}$$
the complex potentials ${\cal Z}$ and ${\cal E}$ satisfy the Ernst equations
$$({\cal Z}+{\cal Z}^{*}) \left [ {\cal Z}_{,tt}-\frac{1}{\omega}(\omega{\cal Z}_{,\omega})_{,\omega} \right ] = 2
\left [ ({\cal Z}_{,t})^2-({\cal Z}_{,\omega})^2 \right ] \eqno{(A3)}$$
and
$$(1-{\cal E}{\cal E}^{*}) \left [ {\cal E}_{,tt}-\frac{1}{\omega}(\omega{\cal E}_{,\omega})_{,\omega} \right ] = 2
\left [({\cal E}_{,t})^2 - ({\cal E}_{,\omega})^2 \right ] \eqno{(A4)} \: .$$
The imaginary part of Eq. (A3) is determined by
$$\left [ \frac{\omega}{\chi^2} q_{2,t} \right ]_{,t} - \left [ \frac{\omega}{\chi^2} q_{2,\omega} \right ]_{,\omega} = 0 \; , \eqno{(A5)}$$
which can be solved, setting
$$\Phi_{,\omega}=\frac{\omega}{\chi^2} q_{2,t} \: , ~~\Phi_{,t}=\frac{\omega}{\chi^2} q_{2,\omega} \: , ~~\Psi = \frac{\omega}{\chi} \; . \eqno{(A6)}$$
Now, introducing the potentials $Z$ and $E$ by 
$$\Psi+i\Phi=Z=\frac{1+E}{1-E} \; , \eqno{(A7)}$$
we verify that $Z$ and $E$ also satisfy Eqs. (A3) and (A4), respectively. Then,
$$\chi=\frac{1-{\cal E}{\cal E}^{*}}{(1-{\cal E})(1-{\cal E}^{*})} \: ,~~q_2=\frac{i({\cal E}^{}-{\cal E})}{(1-{\cal E})(1-{\cal E}^{*})} \eqno{(A8)}$$ and similar expressions can be found also for $\Psi$ and $\Phi$ in terms of $E$. Any solution to the Ernst equations (A3) and (A4) is associated to any of the two distinct complex Ernst potentials ${\cal E}$ or $E$. In this case, having obtained ${\cal E}$ or $E$ by some method, $\nu$ is determined by simple quadratures. Working in terms of ${\cal E}$, we have
$$\nu_{,t}=\frac{\omega}{2\chi^2} \left ( \chi_{,t}\chi_{,\omega} + q_{2,t}q_{2,\omega} \right ) = \frac{\omega}{(1-{\cal E}{\cal E}^{*})^2} \left [ {\cal E}_{,t}{\cal E}^{*}_{,\omega}+{\cal E}^{*}_{,t}{\cal E}_{,\omega} \right ] \eqno{(A9)}$$
and
$$4 \nu_{,\omega} = - \frac{1}{\omega} + \frac{\omega}{\chi^2} \left [ (\chi_{,t})^2 + (\chi_{,\omega})^2 + (q_{2,t})^2 + (q_{2,\omega})^2 \right ] = -\frac{1}{\omega} + \frac{4\omega}{(1-{\cal E}{\cal E}^{*})^2} \left [ {\cal E}_{,t} {\cal E}^{*}_{,t} + {\cal E}^{*}_{,\omega}{\cal E}_{,\omega} \right ] \; . \eqno{(A10)}$$
On the other hand, working in terms of $E$ we obtain
$$ \left ( \nu+\ln{\sqrt{\Psi}} \right )_{,t} = \frac{\omega}{2\Psi^2} \left ( \Psi_{,t}\Psi_{,\omega} + \Phi_{,t} \Phi_{,\omega} \right ) = \frac{\omega}{(1- E E^{*})^2} \left [ E_{,t} E^{*}_{,\omega} + E^{*}_{,t} E_{,\omega} \right ] \eqno{(A11)}$$
and
$$ \left ( \nu + \ln{\sqrt{\Psi}} \right )_{,\omega} = \frac{\omega}{4\Psi^2} \left [ (\Psi_{,t})^2 + (\Psi_{, \omega})^2 + (\Phi_{,t})^2 + (\Phi_{,\omega})^2 \right ] = \frac{\omega}{(1- E E^{*})^2} \left [ E_{,t} E^{*}_{,t}+E^{*}_{,\omega} - E_{,\omega} \right ] \; . \eqno{(A12)}$$
Defining
$$\tilde{\chi}=\frac{\chi}{\chi^2+q^2_2} \: ,~~\tilde{q_2}=\frac{q_2}{\chi^2+q^2_2} \: , ~~\tilde{\Psi} = \frac{\omega}{\chi} \eqno{(A13)}$$
and
$$\tilde{\Phi}_{,\omega}=\frac{\omega}{\tilde{\chi}^2} \tilde{q}_{2,t} \: , ~~\tilde{\Phi}_{,t} = \frac{\omega}{\tilde{\chi}^2} \tilde{q}_{2,\omega} \; , \eqno{(A14)}$$
the complex potentials
$\tilde{{\cal Z}}, \; \tilde{{\cal E}}, \; \tilde{Z}, \; \tilde{E}$, given by the tilde version of Eqs. (A2) and (A7), satisfy Eqs. (A3) and (A4). Thus, any solution to the Ernst equation (A4) does associates to any of the four complex potentials ${\cal E},\tilde{{\cal E}},E$ or $\tilde{E}$. Now, we perform the transformation
$$\omega = \sqrt{\Delta\delta},~~t=\eta\mu \; , \eqno{(A15)}$$
where
$$\Delta = \eta^2+1,~~\delta=\mu^2-1 \; , \eqno{(A16)}$$
while, $\eta$ and $\mu$ range as
$$\eta \in \Re \: ,~~\mu \geq 1 \; , \eqno{(A17)}$$
with $\mu = 1$ corresponding to the axis $\omega = 0$. In this way, we can transfer the problem in the coordinate system $(\eta, \: \mu, \: z, \: \varphi)$, where the metric (A1) reads
$$ds^2 = (\eta^2+\mu^2)e^{2\nu} \left [ \frac{d\eta^2}{\Delta}-\frac{d\mu^2}{\delta} \right ] - \frac{\sqrt{\Delta
\delta}}{\Pi} \left ( dz - q_{2}d\phi \right )^2 - \sqrt{\Delta \delta}\Pi d\phi^2 \; , \eqno{(A18)}$$
since 
$$dt^2 - d\omega^2 = \left ( \eta^2+\mu^2 \right ) \left [ \frac{d\eta^2}{\Delta}-\frac{d\mu^2}{\delta} \right ] \; . \eqno{(A19)}$$
To describe cylindrically-symmetric space times, instead of stationary axisymmetric ones, we perform the analytic continuation
$$\eta \rightarrow i \eta \: ,~~~ p \rightarrow i p \; , \eqno{(A20)}$$
under which, the Ernst equation (A4) reads~\cite{r24},~\cite{r40}
$$({\cal E}{\cal E}^{*}-1) \left [ \Delta ({\cal E}_{,\eta})_{,\eta}-(\delta{\cal E}_{,\mu})_{,\mu} \right ] = 2{\cal
E}^{*} \left [ \Delta ({\cal E}_{,\eta})^2 - \delta ({\cal E}_{,\mu})^2 \right ] \eqno{(A21)}$$ 
and a similar equation is obtained when ${\cal E}\longleftrightarrow E$. Accordingly, for the Kerr-NUT solution (e.g., see~\cite{r24}), we have
$$\Pi=(1-p\eta)^2+(l-q\mu)^2 \: ,~~ Y=p^2\Delta+q^2\delta \: ,~~q^2=p^2+1+l^2 \; , \eqno{(A22)}$$
$$q_{2(\upsilon)}=\frac{2}{pY} \left [ lp^2\Delta(\mu-1)+q(1+l^2-lq-p\eta)\delta \right ] \eqno{(A23)}$$
and
$$\Psi_{(\upsilon)}=\frac{Y}{\Pi} \: , ~~ \chi_{(\upsilon)} = \frac{\omega}{\Psi_{(\upsilon)}} \: , ~~e^{2\nu}_{(\upsilon)} = \frac{\alpha^2 \Pi}{\eta^2+\mu^2} \; . \eqno{(A24)}$$
Upon consideration of the equations
$$\Phi_{,\mu} = \frac{\Delta}{\chi^2}q_{2,\eta} \; ,~~~\Phi_{,\eta}=\frac{\delta}{\chi^2}q_{2,\mu} \; , \eqno{(A25)}$$ we can calculate $\Phi_{(\upsilon)}$ (e.g., see~\cite{r21} -~\cite{r23}), to obtain
$$\Phi_{(\upsilon)}=\frac{2(q\mu-p\eta l)}{\Pi} \; . \eqno{(A26)}$$
Therefore, according to Eq. (A7), the Ernst potential of the Kerr-NUT solution (A18), (A22) - (A24) is
$$ E=\frac{1- i l}{E_k^{*}} \; , \eqno{(A27)}$$
where $E_k=p\eta+ i q\mu $ is the corresponding potential of the Kerr solution.

\begin{acknowledgements}
One of us (D. B. P.) wishes to thank the Department of Fundamental Physics of the University of Barcelona (Spain), where a part of this work took place, for the warm hospitality. The authors would like to express their gratitude to the anonymous referees of this article for their constructive critisism and their comments, which greatly improved the original manuscript.
\end{acknowledgements}



\end{document}